\documentstyle[12pt,psfig]{article}  
\begin{document}
\date{}                  
\bibliographystyle{unsrt}
\def\br{{\bf r}}
\def\DF{\Delta F}
\def\bk{{\bf k}}
\def\bR{{\bf R}}
\def\<{\langle}
\def\>{\rangle}

\title{Design of Force Fields from Data at Finite Temperature}
\author{J.M.Deutsch and Tanya Kurosky \\
University of California, Santa Cruz, U.S.A.}
\maketitle
\abstract
{We investigate the problem of how to obtain the force
field between atoms of an experimentally
determined structure. We show how this problem can be efficiently
solved, even at finite temperature, where the position of the atoms
differs substantially from the ground state.
We apply our method to systems modeling proteins and demonstrate
that the correct potentials can be recovered even in the presence
of thermal noise.}

\vskip 0.3 truein

\section{Introduction}
In many cases it is possible to determine, quite precisely, the structure
of a physical system. X-ray crystallography has made it possible to
determine structures of a myriad of different compounds. Among the
most complicated of these are protein crystals, where thousands of
atoms appear in the unit cell. The structure of many hundreds of proteins
have been determined in this way. The forces between these atoms
are of great importance in predicting the interaction of proteins
with other molecules and also in enabling one to do protein folding
numerically. Therefore there has been a great deal of effort to
determine the forces between sub-molecules in these systems.

One approach has been to determine the forces from {\it ab initio}
quantum calculations of small molecules and additional data obtained from 
experiments on small molecules giving,
for example, resonant frequencies of vibration of
certain bonds. This has led to a number of force fields. For a
review of these, see reference (\cite{MD}). 
These have been used extensively in computational
studies of biological molecules. Such potentials involve many
hundreds of parameters, all of which are quite difficult to
determine. These force-fields are still evolving.

Another approach which is the subject of this letter has been to try to extract
the values of parameters in the force field from the experimentally
determined structures. This approach has some advantages to it
over a direct {\it ab initio} approach. First, the {\it ab initio}
approach has assumed, for the most part, two body potentials and has
ignored higher body terms. At a microscopic level
these other terms should be important. One would like to develop
effective potentials that mimic the higher body terms as well as possible.
By extracting potentials from experimental structures and fitting
them to an effective two body  form,
an optimum two-body force-field which includes higher-body effects
can then in principle be calculated.

Second, the {\it ab initio} approach is intended to describe
the interactions of all the atoms of a protein.  One would like to 
believe, however, that such detail is not necessary in order to predict
the overall structure\cite{Wolynes}.
Coarse grained force fields that consider interactions only between
amino acids can be computed from the experimental structures
\cite{MC,MJ,Hendlich,TD1,TD2,Scheraga,MS}. This may be too crude an approximation for many applications
but has the advantage that it greatly  reduces the complexity
of a protein folding simulation.


One of the most practical approaches along these lines
has been an attempt to derive the energy of interaction of entire amino acids
from their pairing frequency\cite{MJ}. To do so, one treats the protein
as a dilute gas of amino acids, which gives a simple analytical
relation between the pairing frequency and potentials. Despite
the approximate nature of such an approach, this has led to some
success in predicting protein structure\cite{doniach,sun,monge} 
There have been some recent criticisms of the approximations
used\cite{TD1} along with improvements to the method\cite{TD2}.

The purpose of this work is as follows.
We devise and test a method for determining parameters of a
force field from experimental data on molecular structures.
This method finds the set of parameters that will be most likely
to fold the molecules into their observed structures.
Our method is general enough that it can determine the
parameters of a force field of arbitrary complexity, such as the
{\it ab initio} off lattice approaches mentioned above. This method
works correctly even at finite temperature. This is important from
a practical standpoint since the positions of the atoms are only
defined to within a few Angstroms. The problem at finite temperature
is very different than at zero temperature and we will see
that it is a much harder problem. Our solution is very efficient,
and appears on test cases to work remarkably well.

A caveat that we mention is that enough experimental data must be
available in order to determine the correct values of parameters. Even at
finite temperature, we will see that good results can be obtained for 
quite small data sets. It then seems feasible that our method could
be used to determine the force-field of real proteins.

\section{The Problem}

\subsection{Terminology }
Consider a system of $N$ atoms with coordinates $\Gamma \equiv \{\br_i\}$, 
$i = 1,\dots,N$.
The atoms are of different types $s$, and the  chemical sequence
can be denoted $S = \{s_i\}$, $i = 1,\dots,N$. The Hamiltonian for
the system depends on $m$ parameters $P = \{p_i\}$, $i = 1,\dots,m$,
for example, the charge and van der Waals radius. We denote the
Hamiltonian as $H(\Gamma,S,P)$.

The problem is then as follows. Given experimental data on $N_{mol}$ molecules, at
finite temperature, with sequences $S_i$ and configurations ${\Gamma^*}_i$,
what value of parameters $P$ will maximize the probability that  these 
molecules have these experimentally determined structures? 

Very often the parameters can be redefined in such a way
that the Hamiltonian depends on them linearly
\begin{equation}
H(\Gamma, S,P) ~=~ \sum_i^m p_i h_i(\Gamma,S)
\label{eq:linear}
\end{equation}
For example, the van der Waals repulsion between two atoms separated by a
distance $r$ can be written as $K (a/r)^{12}$, where $K$ and $a$ are
parameters. Both of these can be absorbed into a single parameter
$p = Ka^{12}$.

\subsection{Zero Temperature}
If the molecules are in their ground states, then the force on any atom
must be zero. Thus minimizing the sum of the squares of the forces
on all atoms with respect to the parameters $P$, should give a solution
to this problem. Indeed, numerical tests using the model presented in
section \ref{sec:offlattice} confirm that this method works very well
and precisely recovers the values of all parameters, up to an
overall  multiplicative constant. However at any finite temperature
this method fails quite dramatically. In this case the sum of the squares of all
forces can never be chosen to be truly zero. As a result the minimum
is obtained by setting many parameters, such as the charge and
van der Waals radius, equal to zero. At finite temperature, it is
crucial to consider entropic effects and a more fundamental approach
to this problem is required.

For lattice models, the above approach will also not work even at
zero temperature, since the concept of a force is more difficult
to define. For a dense system, it is impossible to make small
displacements, as atoms in the middle of the molecule are already
surrounded by occupied sites. Thus other methods must be employed.

\subsection{The Method}
The formalism used previously to analyze the problem of sequence
design also applies here\cite{us1}. We want to minimize 
\begin{equation}
\DF ~ \equiv ~ \sum_{i=1}^{N_{mol}} ~~ H({\Gamma^*}_i,S_i,P) - F(S_i,P)
\end{equation}
with respect to  the parameters $P$. $\DF$ is the difference between
the energies of the molecules in their experimentally determined
conformations, and their free energies 
\begin{equation}
F(S_i,P) ~=~ -T \ln \sum_{\Gamma} \exp (-\beta H(\Gamma,S_i,P))
\end{equation}
The parameters thus found are optimal in the sense that the molecules
will be more likely to be in their experimentally determined  structures 
$\Gamma_i$ when they interact with these parameters than with any other
choice of parameters.  
The present work attempts to find the solution to a well defined problem.
Other recent  work\cite{MS} chooses a more arbitrary criterion for optimizing
the potential, and will not work at finite temperature. 

In practice however, the calculation of the free energy is a 
formidable task, thus we must devise an efficient method to
minimize $\DF$. 

We start by observing that if we have an approximate solution
$P_0$, the free energy can then be expanded around that point.
For notational simplicity, we will omit the summation over
different molecules, as a single molecule can be redefined
to be composed out of $N_{mol}$ molecules. Corresponding to
the parameters $P_0$, we introduce the Hamiltonian $H_0(\Gamma) \equiv H(\Gamma,S,P_0)$.
\begin{equation}
\DF~\approx~H(\Gamma^*)-F_0-\<H - H_0\>_0+{1\over 2}\beta(\<(H-H_0)^2\>_0-{\<(H-H_0)\>_0}^2)
\label{eq:approx}
\end{equation}
The averages $\<\dots\>_0$ are performed with respect to $H_0$. 
Since $F_0$ is independent of $P$, the minimum of this expression is much easier to determine than that of
the exact one because it involves calculating averages, which is much easier
than calculating free energies. The averaging can be done numerically, say
by molecular dynamics or Monte Carlo. A further simplification can be made
for the class of Hamiltonians that are writable
in the form of (\ref{eq:linear}). In this case  $\DF$ is {\it bi-linear}
in the parameters $P$. That is, it can be written as
\begin{equation}
\DF ~=~ \sum_i^m N_i p_i +{1\over 2}\sum_{i,j=1}^m p_iM_{ij}p_j + {\rm constant}  
\end{equation}
where $N_i$ and $M_{ij}$ are constants that are determined by calculating
the average above.  Because of this, the minimum values of the parameters
can be calculated by solving the matrix equation $M~p ~=~ -N$.

If $P_0$ is not too far from the true minimum, this procedure gives
a better approximation to the minimum of $\DF$ than $P_0$. We
can redefine $P_0$ to be about this new point and then
repeat this procedure iteratively, until the values of parameters
have converged. If $P_0$ is too far, the procedure will not converge,
however we have seen that the radius of convergence is greatly
increased by taking fractional steps in the direction of $P$.
If we regard $P$ as a vector of parameters, then we can take
our new set of parameters to be $\epsilon P + (1-\epsilon)P_0$.

Very interesting recent work\cite{Scheraga} 
using an iterative procedure should 
give similar results at zero temperature.
We do not expect other recent work\cite{MS} to give similar
results even at zero temperature. 

\subsection{Clamping}
Calculating the above averages is still quite difficult because it
involves folding entire molecules with parameters $P_0$, to obtain 
their statistical properties in equilibrium. Even if we start
the molecule off in the experimentally determined conformation
$\Gamma^*$, it will not stay close to there if parameters $P_0$ are quite
different than their true values. Folding real proteins is
still impossible with current computers, so at first sight, the
above method would appear impractical. However, we can circumvent
this problem by adding a $\it clamping$ term to $H_0$.

Folding proteins is difficult because of the many
local minima in the energy landscape, however if we add a clamping
term to the Hamiltonian
\begin{equation}
H_C ~=~ C \sum_i^N |\br_i-\br_i^*|^2
\end{equation}
this localizes the molecule to configurations near the experimentally
determined values $\Gamma^*$. Therefore equilibrating molecules is many
orders of magnitude faster than without this term, even if the value
of $C$ is rather small, allowing the atoms to explore their
local environments. 

So in (\ref{eq:approx}), we add $H_C$ to $H_0$:
\begin{equation}
H_0(\Gamma) ~=~ H(\Gamma_0,S,P)+H_C
\end{equation}
As long as $C$ is small, the second order expansion should still be
a useful approximation.

Once approximate values of  parameters have been determined with
the clamping potential on, it can be gradually turned off. With the
correct parameters for $P_0$, a clamping potential isn't necessary
because the initial configuration we start the molecule in, $\Gamma^*$
is already correctly folded.

This trick works because, unlike the problem of protein folding, we
know the tertiary structure of the molecule and can use that fact to
speed up the averaging.

\section{Application to Lattice Systems}
We apply our method to lattice systems, such as
the HP model. Consider a two dimensional square lattice with
a self avoiding chain interacting with its nearest neighbors. We
assume that there are two species of monomers $\sigma$ that define the
sequence of the chain, of types $\sigma=1$ and $\sigma=2$. 
\begin{equation}
H(\{\sigma_i\}, \{r_i\}) =
{1\over 2} \sum_{i,j}^N V_{\sigma_i\sigma_j}\Delta({\bf r}_i-{\bf r}_j)
\label{eq:model}
\end{equation}     

$\Delta(\br)$ is $1$ if $\br$ is nearest neighbor displacement, and zero
otherwise.
In the HP model\cite{HP},
the interaction between type $i$ and $j$, $V_{ij}$, is especially
simple: $V_{11}=V_{12}=0$ and $V_{22}=-1$. For a given sequence,
the ground state may be degenerate. For $N=14$, there are  386
sequences with unique ground states, so called "good sequences". 
We randomly chose
37 of these ground state sequences as input to our algorithm
which gave predictions for the $V_{ij}$'s\cite{note}. 

We chose $H_0$ to be zero if there was one or more nearest neighbor
contact, and
otherwise, it was infinite. This confines all our averaging to
conformations that have a chance of being a ground state. A conformation
with no contacts cannot be in a unique ground state.
We did not use Monte-Carlo, but instead calculated the averages 
using exact enumeration. This is quite efficient as the averages
in (\ref{eq:approx}) can be written in terms of second and fourth
order correlation functions, $C_{ij} \equiv \<\Delta(\br_i-\br_j)\>$
and $D_{ijkl} \equiv \<\Delta(\br_i-\br_j)\>\<\Delta(\br_k-\br_l)\>$.
These correlation functions are only computed once and so the design
code runs very quickly, over order a few seconds on an Intel 586 machine.

Minimizing (\ref{eq:approx}) gives the values\cite{note} 
$V_{11} = 0.057,V_{12} = 0.14$, and $V_{22} = -1$. This might
seem to be quite far off from the original values, however
refolding the 37 chains using these new values gives precisely
the same ground states for all the chains. In other words this
potential gives the same ground state as the original.

For a commonly used variant\cite{SG1,gap} of the Dill and Lau model, 
there are 1619 good sequences. In this case, the values found are\cite{note}
$V_{11} = -0.89,V_{12} = .28$, and $V_{22} = -1$. Again, this correctly
refolds all 37 conformations considered to the correct ground states.

In both cases, the method reproduces the correct ground states immediately,
so that  an iterative method need not be considered. We now turn to a
continuous system at finite temperature. 

\section{Application to an Off-Lattice System}
\label{sec:offlattice}
We now consider an off-lattice system containing much of the essential
physics of a real protein. We consider a system of atoms connected
by springs with an equilibrium length $r_0$, and spring coefficient
$k$. We also say that there are two types of atoms with charge $q_i$ of
either $Q$ or $-Q$. Finally we include a Van der Waals repulsion $(a/r)^{12}$. 
The Hamiltonian is then
\begin{equation}
H ~=~ \sum_i^N {k\over 2}(r_i-r_0)^2 + \sum_{i<j}^N {q_iq_j\over |\br_i-\br_j|} + 
      ({a \over |\br_i-\br_j|})^{12}
\end{equation}
which depends on the parameters $k$, $r_0$, $Q$, and $a$. This Hamiltonian
can be rewritten in the form of (\ref{eq:linear}).

\begin{figure}[tbh]
\begin{center}
\                
\psfig{file=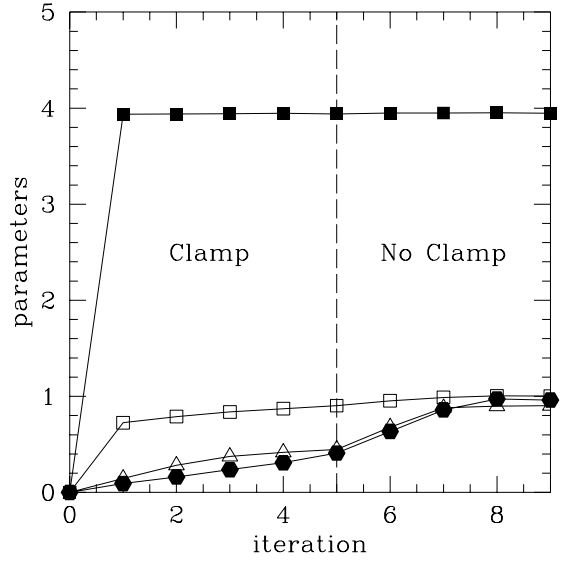,width=3in}
\end{center}
\caption{The computed  values of the parameters as a function of the number 
of iterations for the off-lattice model considered in the text. The 
spring constant $k$ is denoted by the open triangles. 
The equilibrium spring length $r_0$ by solid squares,
the charge $Q$ by solid hexagons, and the van der Waals radius $a$ by open
squares}
\label{fig:params(t)}
\end{figure}

To test our method, we first made a database of  12 structures, with
12 different sequences of the $q_i$. We chose some fixed  values
for the parameters, $k = 1$, $r_0 = 4$, $Q = 1$ and $a = 1$.
We cooled the atoms using simulated annealing down to
a temperature where they had collapsed to well defined structures,
$\beta = 20$. Then we fed these structures into our program, which uses Monte-Carlo
to estimate the averages in (\ref{eq:approx}). We applied a moderate
clamping potential with $C= 2.5$ for 5 iterations and then turned it
off and continued to iterate 4 more times. The program is suppose to
determine the parameters $k$, $r_0$, $Q$, and $a$ from only the database
of these twelve structures.  The results are displayed in
figure \ref{fig:params(t)}. The results took about five minutes on
an Intel 586 microprocessor. The computed parameters are within 12\%
of the real values.

\section{Conclusions}
We have presented a new and relatively simple method for determining
forces between atoms from their structure at finite temperature.
We have applied this to several model systems, on lattice and
off lattice, and have found that it gives accurate results very
efficiently. Our approach expands $\DF$ introduced earlier\cite{us1}
to second order about some approximate parameters. $\DF$ is again
minimized, and the procedure is repeated iteratively until satisfactory
convergence is obtained.  Because of the efficiency of this method,
it appears computationally feasible to us to apply our method to
real protein data bases. This is currently under investigation.

\section{Acknowledgements}
We wish to thank Douglas Williams for useful discussions.
This work is supported by NSF grant number DMR-9419362
and acknowledgment
is made to the Donors of the Petroleum Research Fund, administered
by the American Chemical Society for partial support of this research.

\newpage                               

\begin{thebibliography}{99}
\bibitem{note} Note that because
we are feeding in states at zero temperature, the potential
is only determined to within an overall multiplicative factor.
\end{thebibliography}
\end{document}